\documentstyle[12pt]{article}

\author{ E.V.\ Damaskinsky \\
{\em Institute of Military Constructing Engineering,} \\
{\em Zacharievskaya street 22,} \\
{191194, Sanct Petersburg, Russia}
\and \underline{M.A.\ Sokolov} \\
{\em Sanct Petersburg Institute of Mashine Building,} \\
{\em Poliustrovskii prospect 14,}\\
{195108, Sanct Petersburg, Russia} \\
{e-mail derzh@pmash.spb.su} }
\title{\bf Some remarks \\
on the Gauss decomposition \\
for quantum group $ GL_q(n) $ \\
with application to q-bosonization }
\date{\today}
\newcommand{\beq}{\begin{equation}}
\newcommand{\eeq}{\end{equation}}
\newcommand{\beeq}{\begin{eqnarray}}
\newcommand{\eeeq}{\end{eqnarray}}
\newcommand{\bdm}{\begin{displaymath}}
\newcommand{\edm}{\end{displaymath}}
\newcommand{\gl}[1]{{$GL_q$}$(#1)$}

\newtheorem{PR}{Proposition}
\input tcilatex
\QQQ{Language}{American English}

\begin{document}

\maketitle

\begin{abstract}

In this letter some properties of the Gauss decomposition of quantum
group $GL_q(n)$ with application to q-bosonization are considered.

\end{abstract}

\newpage

For investigations of algebraic systems it is often useful to handle with
theirs homomorphic realizations by means of simpler systems. The
 representation of Lie algebra generators by creation and annihilation
operators of bosonic oscillators (bosonization) supplies us with a well
known example. When we turn to more complicated case of quantum groups and
quantum algebras [1] it is natural to change the usual oscillators to
deformed ones (q-oscillators) [2-6]. The q-bosonization procedure for
quantum algebras was given in [5] with sufficient completeness.  However,
the situation for quantum groups is quite different. It seems first attempts
to this end have been done for the quantum group $GL_q(n)$ in the special
case $q^n=1$ only [7-9]. In the general case, $q \in C \backslash \{0\}$,
the examples of q-bosonization were given in [10] and [11] for the $GL_q(2)$
and $GL_q(3)$ respectively.  The attempt to generalize these results
to the $GL_q(n)$  was undertaken in somewhat complicated fashion in Ref. [12].
The deficiency of this approach [10,12] consists in the use of specific
features of the Fock representation for the q-oscillators. As a result,
 a wide
class of non Fock representations (listed, for instance, in [6,13-14]) are
excluded from considerations. We must also mention the interesting works
[15-16] concerned with the similar problem for the matrix pseudogroups
$S_\mu U(n)$ and used somewhat different form of q-oscillators.

In previous paper [17] we showed that the Gauss decomposition for the
$GL_q(2)$ suggests q-bosonization in a very simple and pure algebraic way.
In this letter we shall give some remarks on general properties of the
$GL_q(n)$ Gauss decomposition and generalize to this quantum group the
algebraic procedure of q-bosonization suggested earlier [17] for the
$GL_q(2)$. As an illustration of the results we shall consider the $GL_q(3)$
case in some details.

   2. In this section we shall briefly remind some definitions and results
of Ref.[1]. The quantum group \gl{n} is defined as an associative
unital C-algebra which is generated by $n^2$ elements $T_{ij}$ subject to
the commutation relations
\beq
\begin{array}{cccccc}
 T_{ij}T_{ik} &=& qT_{ik}T_{ij},& \hspace{.7cm} T_{ik}T_{lj} &=&\
 T_{lj}T_{ik},\\ T_{ik}T_{lk} &=& qT_{lk}T_{ik},&\hspace{.7cm}
 \mbox{$[T_{ij},T_{lk}]$} & = & \mbox{$\lambda T_{ik}T_{lj}$},
\end{array}
\eeq
where
\( q \in C \backslash \{0\},\  \lambda = q-q^{-1} ,\
\ 1 \leq j < k \leq n,\  1 \leq i < l \leq n \).
In addition,
the $ GL_q(n) $ is endowed with a Hopf
algebra structure which introduced by the three maps: a
comultiplication $\Delta$ , a counit $\epsilon$ and  an  antipod  $S$  [1].

Let $T = (T_{ij})$ be $n$ by $n$ matrix of $GL_q(n)$ generators (q-matrix).
{}From the commutation relations (1) it follows that the quantum
(q-)determinant
\beq
D_q(T) \equiv det_qT = \sum_{\sigma}(-q)^{l(\sigma)}
    \prod_{i=1}^{n} T_{i \sigma (i)}
\eeq
belongs to the center of the Hopf algebra $GL_q(n)$ . In the definition (2)
the sum is over all the permutations $\sigma$ of the set $ (1,2,...,n) $ ,
and $\sigma$ ($l$) is length of $\sigma$ . There is additional assumption:
\( det_qT \neq 0 \) . Supposing the invertibility of $D_q(T)$ , one can
calculate the two-sided inverse matrix $T^{-1}$
\bdm (T^{-1})_{ik} =
           (D_q(T))^{-1} (-q)^{i-k} D_q(M_{ik}),
\edm
where $D_q(M_{ik})$ is a q-determinant of the minor matrix $M_{ik}$
which was obtained from $T$ by removing of the $i$ -th row and $k$ -th
column.  In the case of $GL_q(n)$ the above mentioned coalgebra maps are
specified by the formulas
\beq \Delta (T) = T ( \otimes ) T, \hspace{5mm}
      \epsilon (T) = 1 ,      \hspace{5mm}
      S(T) = T^{-1},
\eeq
where 1 is the unit matrix and $( \otimes )$ refers to the usual
matrix multiplication with a tensor product of matrix elements.

In R-matrix approach [1] the equation
\beq
    RT_1T_2=T_2T_1R,
\eeq
encodes commutation relations between generators of a quantum group.
In this equation $R$ is a square number matrix of order $n^2$ , $T$
is a q-matrix and $T_1=T( \otimes )1, T_2=1( \otimes )T$ .
The commutation relations (1) follow from the eq.(4) for the $R$ -matrix
corresponding to the Lie algebra $sl(n)$ .  The Lie group $GL(n)$
can be defined as an endomorphism group of complex $n$ -dimensional
linear space $C^n$ . In complete analogy, the quantum group \gl{n} ,
can be defined by its co-action $\delta$ on the related quantum space
$C_q^n$ .  This view gives the base of non $R$ -matrix approach to
quantum deformations [18-20].

3. Let us consider the Gauss decomposition for the \gl{n} $q$ -matrix
$T$ ,
    \[ T=(T_{ik})=T_LT_DT_R,\]
where $T_L=(u_{ik})$ and $T_R=(z_{ik})$ are upper- and lower-triangular
matrices respectively with units at theirs diagonals, and $T_D=(A_{ik})$,
with $A_{ik}= \delta _{ik}A_{k}$, is a diagonal matrix.  It is not
dificult to express "old" generators $T_{ik}$ in terms of the "new" ones,
that is by the elements of the matrices $T_L, T_D$ and $T_R$.  For
convinience we shall formulate the main properties of the Gauss
decomposition in the following propositions.
\begin{PR} The commutation
relations (1) of the quantum group $GL_q(n)$ are fulfilled if the following
commutation relations for the elements of the $T_L, T_D$ and $T_R$  matrices
hold
\begin{eqnarray} A_{k}u_{ij}=q^{\delta _{ik}-\delta _{jk}}u_{ij}A_k, &
		       \hspace{15mm}1 \leq i,j,k \leq n; \nonumber  \\
 u_{ik}u_{jk}=q^{\delta _{ik}-\delta _{jk}+1}u_{jk}u_{ik}, &
                     \hspace{15mm} 1 \leq k \leq n,\   i<j; \nonumber \\
 u_{ik}u_{il}=q^{\delta _{ij}-\delta _{ik}+1}u_{ij}u_{ik}, &
                     \hspace{15mm}  1 \leq i \leq n, \   k<j; \nonumber \\
 u_{ik}u_{lj}=q^{\delta _{ij}-\delta _{kl}}u_{lj}u_{ik}, &
                     \hspace{15mm}   i<l, \   k>j; \nonumber \\
 u_{ik}u_{lj}-q^{\delta _{ij}-\delta _{kl}}u_{lj}u_{ik}= & \nonumber \\
       = \lambda q^{\delta _{ik}-\delta _{kl}}u_{lk}u_{ij}, &
                     \hspace{15mm}       i<l, \  k<j; \nonumber \\
 \left[ u_{ik}, z_{jl}\right] =[A_i,A_k]=0, & \hspace{15mm}1 \leq i,j,k,l
\leq n; \\ A_{k}z_{ij}=q^{\delta _{kj}-\delta _{ik}}z_{ij}A_k, &
\hspace{15mm} 1 \leq i,j,k \leq n; \nonumber \\ z_{ik}z_{jk}=q^{\delta
_{jk}-\delta _{ik}+1}z_{jk}z_{ik}, & \hspace{15mm} 1 \leq k \leq n, \ i<j;
 \nonumber \\ z_{ik}z_{ij}=q^{\delta _{ik}-\delta _{ij}+1}z_{ij}z_{ik}, &
 \hspace{15mm}  1 \leq i \leq n, \  k<j; \nonumber \\ z_{ik}z_{lj}=q^{\delta
		     _{kl}-\delta _{ij}}z_{lj}z_{ik}, & \hspace{15mm}i<l, \
 k>j; \nonumber \\ z_{ik}z_{lj}-q^{\delta _{kl}-\delta _{ij}}z_{lj}z_{ik}= &
 \nonumber \\ = \lambda q^{\delta _{kl}-\delta _{ik}}z_{lk}z_{ij}, &
		     \hspace{15mm}i<l, \  k<j; \nonumber
\end{eqnarray}
\end{PR}

\begin{PR}
      In terms of
     the "new" generators the quantum determinant $D_q(T) = det_qT$ has the
     simple form
\beq    D_q(T) = detT_D=A_{11}A_{22} \cdot \ldots \cdot A_{nn}   \eeq
    and commutes with every element of the matrices $T_L, T_D$ and $T_R$.
\end{PR}

(For the \gl{2} case the expression (6) was given in [19]). These two
     propositions can be checked by direct calculations.

Let us denote \( \widetilde{T} =T_DT_R, \text{\ and\ } \widehat{T}=T_LT_D \)
.  We would like to stress that the elements each of the matrices \( (T_L,
T_D, T_R, \widetilde{T}, \text{\ and\ } \widehat{T})  \) form a set closed
under the commutation relations (5) and, thus, define a deformed algebra.

\begin{PR}
{\em a)} The commutation relations between the elements of the matrix
  \( \widetilde{T} ( \text{\ and\ }  \widehat{T} ) \)
are determined by the quantum group equation {\em (4)} with the
same $R$-matrix as for \gl{n};\\ {\em b)} There is no $R$-matrix that gives
the commutation relations between the elements of the matrix $T_R$ (and
  $T_L$).  \end{PR}

The first part of this proposition can be proved by direct verification.
The attempts of calculation of appropriate $R$-matrix from the equation (4)
in the case of the matrices $T_R$ and $T_L$ inevitably leads to
contradictions.  So the algebras defined by the elements of matrices  $T_R$
and $T_L$, supply us with examples of non $R$-matrix quantum deformations.

Using the homomorphic property of a  comultiplication in  the \gl{n} we can
define the comultiplications in the algebras connected with each of the
matrices \( T_R, T_D, T_L, \widetilde{T} \text{\ and\ }\widehat{T} \). (For
the \gl{2}  this was done in $[21]$). However such inherited
comultiplication has cumbersome form even for $n=2,3$.

\begin{PR}
The algebra generated by the elements of the matrix \(\widetilde{T}\)
is a Hopf algebra {\em(}under the co-operations defined in {\em(3))} and
gives us an example of a new quantum group.  The same is true for
\(\widehat{T}\).
\end{PR}

\begin{PR}
The map
\[ \delta (X)=T(\otimes )X, \hspace{10mm} \delta (x_i)=
\sum_{k=1}^{n} T_{ik} \otimes x_k \]
defines the co-action of the quantum
groups \(\widetilde{T}\) and $\widehat{T}$ on  the  quantum space $C_q^n$.
Moreover, the maps  $\widetilde{\delta}$ : $ C_q^n \rightarrow \widetilde{T}
(\otimes ) C_q^n$  and  $\widehat{\delta}$ : $ C_q^n \rightarrow \widehat{T}
(\otimes ) C_q^n$  are algebra homomorphismes and endow the $C_q^n$ with
a left $\widetilde{T}$-  and $\widehat{T}$-comodule structure,
respectively.  \end{PR}

The proof of these two propositions is evident in  view  of  the
proposition 3.a).

4.  Let \{ $a_i ^{\dag}, a_i, N_i$\} and \{ $b_i ^{\dag}, b_i, M_i$\} ,
\hspace{5mm} $i,j=1 \div n $ be  two independent families of mutually
commuting $q^{-1}$- and $q$-oscillators [2-6] defined by the relations
\beq
 a_ia_i^{\dag} -q^{-1}a_i^{\dag} a_i=q^{N_i} , \hspace{5mm} N_ia_i ^{\dag} =
 a_i ^{\dag} (N_i+1), \hspace{5mm} N_ia_i =a_i(N_i-1);
\eeq \beq
 b_ib_i^{\dag} -qb_i^{\dag} b_i=q^{-M_i} , \hspace{5mm} M_ib_i ^{\dag} =
 b_i ^{\dag} (M_i+1) ,\hspace{5mm} M_ib_i =b_i(M_i-1);
\eeq
Define
\beq
u_{ik}=\left\{ \begin{array}{ll}
	                 0              & if \  i>k \\
                         1              & if \  i=k \\
        f_{ik}q^{N_{ik}}a_i ^{\dag} a_k & if \  i<k
              \end{array} \right. ; \hspace{5mm}
z_{ik}=\left\{ \begin{array}{ll}
	                 0              & if \  i<k \\
                         1              & if \  i=k \\
       g_{ik}q^{-M_{ik}}b_i ^{\dag} b_k & if \  i>k
       ; \hspace{5mm}
               \end{array}\right.
\eeq

\beq A_{ik}=\delta _{ik} q^{N_{k}-M_{k}}; \eeq
where
\[ N_{ik}=\sum_{j=i+1}^{k-1}N_{j}, \hspace{5mm}
   M_{ik}=\sum_{j=k+1}^{i-1}M_{j},\]
and \(N_{ik}=0\  (M_{ik}=0)\) if \( i>(k-2), \ (i<(k+2)) \).  Then we have
\[ det_qT =
\prod_{i=1}^{n} A_{ii} = q^{N-M}, \hspace{5mm} \ N = \sum_{j=1}^{n}N_{j},
\hspace{5mm} M = \sum_{j=1}^{n}M_{j}. \]

\begin{PR}
The expressions {\em (9-10)} satisfy the commutation relations {\em (5)} if
number coefficients $f_{ik}$ and $g_{ik}$ fulfil the equations
\beeq
 \left\{ \begin{array}{lcll}
     f_{ij}f_{kl}\! &\! =\! &\! f_{kj}f_{il} &  i\!<\!k\!<\!j\!<\!l \\
     f_{ij}f_{jk}\! &\! =\! &\! \lambda q^{-1}f_{ik} &  i\!<\!j\!<\!k
         \end{array} \right.\!\! ;\! &
 \left\{ \begin{array}{lcll}
     g_{ij}g_{kl}\! &\! =\! &\! g_{kj}g_{il}  &  k\!>\!i\!>\!l\!>\!j \\
     g_{ij}g_{ki}\! &\! =\! &\! \mbox{}-q\lambda g_{kj}  &  k\!>\!j>\!i
         \end{array} \right.\!\! ;
\eeeq
In this case the formulas {\em (9-10)} realize q-bosonization of the quantum
groups $T_R,\ T_D,\ T_L$ and, consequently $\widetilde{T},\ \widehat{T},\
GL_q(n).$
\end{PR}

There are several solutions of the equations (11). The  symplest one is
\beq  f_{ij} = \lambda /q, \hspace{5mm}  g_{ij} = -q\lambda.
\eeq

It is  worth  noting  that  in  the  present  version  of $GL_q(n)$
q-bosonization 2n independent deformed oscillators are used.  On  the
other hand, following the method  of  Refs.[10-12]  for  this  purpose
n(n-1)/2 q-oscillators are necessary.

5. To illustrate the above results, let us consider  the  quantum
group $GL_q(3)$ as an example. The case of $GL_q(2)$ studied in [17], is
not interesting enough because both the matrices $T_R$ and $T_L$  contein
one non-trivial element only. The Gauss decomposition for the \gl{3}
has the form
\bdm
T=T_LT_DT_R=\left( \begin{array}{ccc}
                      1 & u & v \\
                      0 & 1 & w \\
                      0 & 0 & 1
             \end{array} \right)
	    \left( \begin{array}{ccc}
                      A & 0 & 0 \\
                      0 & B & 0 \\
                      0 & 0 & C
            \end{array} \right)
	    \left(\begin{array}{ccc}
                      1 & 0 & 0 \\
                      x & 1 & 0 \\
                      y & z & 1
            \end{array} \right) =
\edm
\bdm
            \left( \begin{array}{ccr}
                  A+uBx+vCy & uB+vCz & vC \\
                    Bx+wCy  &  B+wCz & wC \\
                      Cy    &   Cz   &  C
            \end{array} \right),
\edm
\bdm
\widetilde{T}=T_DT_R=\left( \begin{array} {llc}
                         A  & 0  & 0 \\
                         Bx & B  & 0 \\
                         Cy & Cz & C
                     \end{array} \right) , \hspace{5mm}
\widehat{T}=T_LT_D= \left( \begin{array}{crr}
		         A & uB & vC \\
		         0 &  B & wC \\
		         0 &  0 &  C
                     \end{array} \right).
\edm

The commutation relations for "new" generators follow from (5)
\bdm
\begin{array}{c}
    $ [{\em x,u}]  =  [{\em x,v}]  =  [{\em x,w}]  = 0,$ \\
    $ [{\em y,u}]  =  [{\em y,v}]  =  [{\em y,w}]  = 0,$ \\
    $ [{\em z,u}]  =  [{\em z,v}]  =  [{\em z,w}]  = 0,$ \\
    $ [{\em A,B}]  =  [{\em A,C}]  =  [{\em B,C}]  = 0;$
\end{array}
\edm
\bdm
\begin{array}{ccc}
   vw=qwv,\hspace{7mm} & uv=qvu, \hspace{7mm} & quw-wu=\lambda v, \\
   xy=qyx,\hspace{7mm} &yz=qzy, \hspace{7mm}  &xz-q^{-1}zx=\lambda y;
\end{array}
\edm
\beq
\begin{array}{ccc}
   uB=qBu, \hspace{7mm} & uC=Cu, \hspace{7mm} & Au=quA, \\
   vC=qCv, \hspace{7mm} & vB=Bv, \hspace{7mm} & Av=qvA, \\
   wC=qCw, \hspace{7mm} & wA=wA, \hspace{7mm} & Bw=qwB;
\end{array}
\eeq
\bdm
\begin{array}{ccc}
   xB=qBx, \hspace{7mm} & xC=Cx, \hspace{7mm}  & Ax=qxA, \\
   yC=qCy, \hspace{7mm} & yB=By, \hspace{7mm}  & Ay=qyA, \\
   zC=qCz, \hspace{7mm} & zA=Az, \hspace{7mm}  & Bz=qzB.
\end{array}
\edm
The quantum determinant $D_q(T) = ABC$ commutes with all of "new" generators
and, hence, with $\widetilde{T}$ and $\widehat{T}$ .

Applying the usual comultiplication \( \Delta(T)=T( \otimes ) T \) in the
\gl{3} and denoting
  \[ Q=y \otimes v + z \otimes w + 1 \otimes 1 \equiv y_1v_2+z_1w_2+1; \]
  \[ E = 1+x_1u_2-(x_1v_2+w_2)Q^{-1}(y_1u_2+z_1) ; \]
  \[ K = (1-u_2E^{-1}L-v_2Q^{-1}[y_1-(y_1u_2+z_1)E^{-1}L]), \]
where $L=[x_1-(x_1v_2+w_2)Q^{-1}y_1]$, one can obtain for inheritable
comultiplication the following expressions
\[  \Delta (A) = A_1KA_2, \hspace{7mm} \Delta(B)=B_1EB_2, \hspace{7mm}
    \Delta (C) = C_1QC_2 \]
\[  \Delta (u) = u_1+
     A_1u_2(B_1E)^{-1}-A_1v_2Q^{-1}(y_1u_2+z_1)(B_1E)^{-1}, \]
\[  \Delta (v) = v_1+ (A_1v_2+u_1B_1(x_1v_2+w_2))Q^{-1}C_1^{-1}, \]
\[  \Delta (w)= w_1+B_1(x_1v_2+w_2)Q^{-1}C_1^{-1}, \]
\[  \Delta (x)= x_2+ (EB_2)^{-1} \lbrack x_1-(x_1v_2+w_2)Q^{-1}y_1
     \rbrack A_2, \]
\[  \Delta (y)=y_2+C_2^{-1}Q^{-1} \lbrack y_1A_2+(y_1u_2+z_1)B_2x_2
    \rbrack , \]
\[  \Delta (z)=C_2^{-1}Q^{-1}(y_1u_2+z_1)B_2+z_2. \]
We remaind that $A_1=A \otimes 1$, $A_2=1 \otimes A$, etc. As it was pointed
above, there is "natural" comultiplication together with "inheritable" one
in $\widetilde{T}$ and $\widehat{T}$. These two maps are not equivalent
because they give obviously different expression for the element $C=T_{33}$.
Therefore we deal with  new quantum groups.

For the quantum group $\widetilde{T}$ the natural comultiplication is given
by the relation
\bdm
\Delta (\widetilde{T})=\left( \begin{array}{llc}
                                            A  & 0  & 0 \\
                                            Bx & B  & 0 \\
                                            Cx & Cz & C
                              \end{array}\right) ( \otimes )
\left(\begin{array}{llc} A &0 &0 \\ Bx &B &0 \\ Cy &Cz &C
\end{array} \right) =
\edm
\bdm
=\left(\begin{array}{ccc}  A \otimes A &0 &0 \\
Bx \otimes A +B \otimes Bx &B \otimes B &0 \\
Cy \otimes A +Cz \otimes Bx +C \otimes C &Cz \otimes B+C \otimes Cz
  &C \otimes C
\end{array} \right),
\edm
which yields
\beq  \Delta (A)= A \otimes A ; \hspace{5mm} \Delta (B)= B \otimes B ;
      \hspace{5mm} \Delta (C) = C \otimes C ;
\eeq
\beq
\begin{array}{c}
 \Delta (x)= x \otimes B^{-1}A+1 \otimes x,  \\
 \Delta (y)= y \otimes C^{-1}A+z \otimes C^{-1}B+1 \otimes y, \\
       \Delta (z)= z \otimes C^{-1}B+1 \otimes z.
\end{array}
\eeq
(we assume that the elements $A, B, C$ are invertible). In  close
analogy, for the quantum group $\widehat{T}$ besides (13) one has
\beq
\begin{array}{c}
  \Delta (u) = AB^{-1} \otimes u+ u \otimes 1 ,\\
  \Delta (v) = AC^{-1} \otimes v+uBC^{-1} \otimes w+ w \otimes 1,\\
  \Delta (w) = BC^{-1} \otimes w+ w \otimes 1.
\end{array}
\eeq
It is not dificult to check that the matrices \bdm
 \widetilde{T}^{-1}=\left(\begin{array}{ccc}
 A^{-1} &0 &0 \\ -qA^{-1}x &B^{-1} &0 \\ q^2A^{-1}(xz-qy) &-qB^{-1}z &C^{-1}
\end{array} \right) ;\edm \bdm
\widehat{T}^{-1}=\left(\begin{array}{ccc}
  A^{-1} &-quA^{-1} &q^{-1}(uw-v)A^{-1} \\
  0      &B^{-1}    &-q^{-1}wB^{-1} \\
  0      &0         &C^{-1}  \end{array} \right), \edm
are two sided inverse ones to the $\widetilde{T}$ and $\widehat{T} $
respectively.  Therefore, the antipodes and counits can be determined in
concordance  with (3).  On  element level one has
\[ S(A)=A^{-1}, \hspace{7mm} S(B)=B^{-1}, \hspace{7mm} S(C)=C^{-1},\]

\[S(x)=-q^{2}A^{-1}Bx=-xA^{-1}B,\hspace{5mm}
                S(z)=-q^{2}B^{-1}Cz=-zB^{-1}C,\]
\[ S(y)=q^{3}A^{-1}C(xz-qy)=(zx-y)A^{-1}C;\]

\[S(u)=-q^{-1}BuA^{-1}=-BA^{-1}u,\hspace{5mm}
                S(w)=-q^{-1}CwB^{-1}=-CB^{-1}w,\]
\[ S(v)=q^{-1}C(uw-v)A^{-1}=CA^{-1}(uw-v);\]

\[ \varepsilon (A) =\varepsilon (B) =\varepsilon (C) =1,\hspace{3mm}
   \varepsilon (u) =\varepsilon (v) =\varepsilon (w) =0,\hspace{3mm}
   \varepsilon (x) =\varepsilon (y) =\varepsilon (z) =0.\]
All the Hopf algebra axioms [1] with these determinations are satisfied. As
a result, the algebras associated with $\widetilde{T}$ and $\widehat{T}$
matrices  are  endowed with a Hopf algebra structure. Therefore, they indeed
can  be  considered as new quantum groups.

6. Now we turn to q-bosonization of the \gl{3} using  the  solution
(12) of the equations (11). So taking three $ q^{-1} $-oscillators (7) and
three $q$-oscillators (8) and using (9-10) one gets
\[u=\lambda q^{-1}a_1^{\dag}a_2,\hspace{5mm}
v=\lambda q^{-1}q^{N_{2}}a_{1}^{\dag}a_{3}, \hspace{5mm}
w=\lambda q^{-1}a_{2}^{\dag}a_{3},\]
\[x=-\lambda qb_{2}^{\dag}b_{1},\hspace{5mm}
y=-\lambda qq^{-M_{2}}b_{3}^{\dag}b_{1},\hspace{5mm}
z=-\lambda qb_{3}^{\dag}b_{2},\]
\[ A= q^{N_{1}-M_{1}},\hspace{5mm} B= q^{N_{2}-M_{2}},   \hspace{5mm}
 A= q^{N_{3}-M_{3}}.\]
For the original \gl{3}-generators this gives
\[ T_{11}= q^{N_{1}-M_{1}}-q\lambda ^{2} q^{N_{2}-M_{2}}\left[
         a_{1}^{\dag}a_{2}b_{2}^{\dag}b_{1}+ q^{N_{3}-M_{3}}
         a_{1}^{\dag}a_{3}b_{3}^{\dag}b_{1}\right] ,\]
\[ T_{12}=\lambda q^{-1} \left[a_{1}^{\dag}a_{2}
          q^{N_{2}-M_{2}}-\lambda qq^{N_{2}+N_{3}-M_{3}}
          a_{1}^{\dag}a_{3}b_{3}^{\dag}b_{2}\right] ,\]
\[ T_{13}=\lambda q^{-1} a_{1}^{\dag}a_{3}q^{N_{2}+N_{3}-M_{3}},\]
\[ T_{21}=-\lambda q \left[ q^{N_{2}-M_{2}}b_{2}^{\dag}b_{1}+\lambda
          q^{N_{3}-M_{3}-M_{2}}a_{2}^{\dag}a_{3}b_{3}^{\dag}b_{1} \right],\]
\[ T_{22}=q^{N_{2}-M_{2}} - \lambda  ^{2}q^{N_{3}-M_{3}}a_{2}^{\dag}a_{3}
           b_{3}^{\dag}b_{2},\]
\[ T_{23}=\lambda q^{N_{3}-M_{3}}a_{2}^{\dag}a_{3} , \]
\[ T_{31}=-\lambda q q^{N_{3}-M_{2}-M_{3}}b_{3}^{\dag}a_{1},\]
\[ T_{32}=-\lambda q q^{N_{3}-M_{3}}b_{3}^{\dag}b_{2} ,\]
\[ T_{33}= q^{N_{3}-M_{3}}.\]

\vspace{1.5cm}
The authors thanks Prof. P.P.Kulish for useful discussions. The work of
one of us (D.E.V.) was supported in part by the Soros Foundation
(individual grant for 1993) and Russian Fond of Fundamental
    Researches (grant No 94-01-01157-a).

\end {document}